\pdfoutput=1
\PassOptionsToPackage{numbers,sort&compress}{natbib}
\PassOptionsToPackage{colorlinks=true,allcolors=blue}{hyperref}

\documentclass[reprint,aps,prapplied]{revtex4-2} 

\usepackage{graphicx}
\usepackage{amsmath,amssymb,mathtools,bm}
\usepackage{booktabs}
\usepackage{float}

\usepackage{dcolumn}
\usepackage{verbatim}
\usepackage{xcolor}
\usepackage[most]{tcolorbox}
\usepackage{soul}
\usepackage[framemethod=TikZ]{mdframed}
\usepackage{tikz}

\usepackage{natbib}
\usepackage{hyperref} 

\begin{document}

\title{Dynamic Josephson Junction Metasurfaces for Multiplexed Control of Superconducting Qubits}

\author{Mustafa Bakr}%
\email{mustafa.bakr@physics.ox.ac.uk} 
\affiliation{Department of Physics, Clarendon Laboratory, University of Oxford, Oxford OX1 3PU, United Kingdom
}

\begin{abstract}
Scaling superconducting quantum processors to large qubit counts faces challenges in control signal delivery, thermal management, and hardware complexity, particularly in achieving microwave signal multiplexing and long-distance quantum information routing at millikelvin (mK) temperatures. We propose a space-time modulated Josephson Junction (JJ) metasurface architecture to generate and multiplex microwave control signals directly at mK temperatures. Theoretical and numerical results demonstrate the generation of multiple frequency tones with controlled parameters, enabling efficient and scalable qubit control while minimizing thermal loads and wiring overhead. We derive the nonlinear wave equation governing this system, simulate beam steering and frequency conversion, and discuss the feasibility of experimental implementation. These results lay the groundwork for a next-generation cryogenic signal-delivery paradigm that may enable scaling superconducting quantum processors to thousands of qubits without overwhelming limited dilution-refrigerator cooling power.
\end{abstract}

\maketitle


\section{Introduction}
Solid-state quantum processors such as superconducting qubits \cite{ref0, ref1, ref2, ref3, ref4, ref5, ref6, ref7, ref8, ref9, ref10, ref11, ref12, ref13, ref14, ref15} and semiconductor quantum dots \cite{ref16} operate at millikelvin temperatures to initialise the system in their ground state and prevent thermal excitation during operation. Initialisation, control, and measurement are routinely executed using classical control electronics generating microwave pulses \cite{ref17, ref18}. Scaling these approaches from the few-qubit level to large-scale quantum processors introduces challenges beyond just space limitations \cite{ref19, ref20, ref21, ref22, ref23, ref24, ref25, ref26, ref27, ref28, ref29}. Significant passive and active heat loads due to the dissipation of control signals in cables and attenuators are major concerns. These issues become more pronounced as scaling requires an increasing number of microwave and DC cables to be integrated into the dilution refrigerator (DL). 

\begin{figure*}[!t]
	\begin{center}		
    \includegraphics[width=1.5\columnwidth]{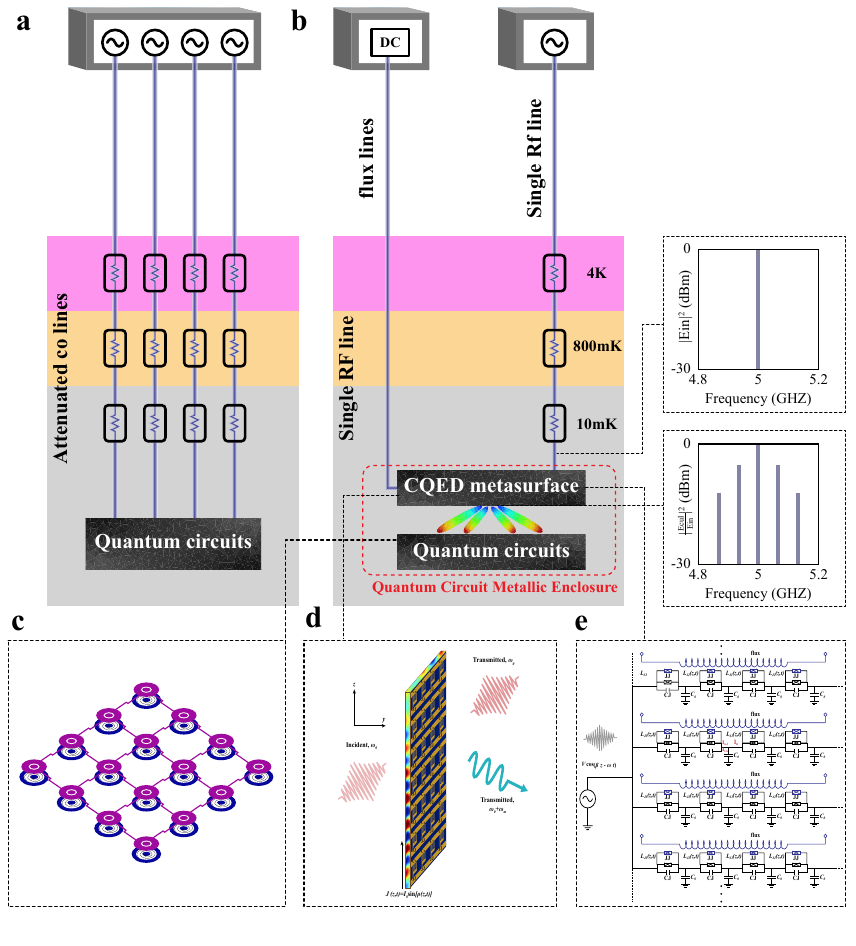}
	\caption{a) Schematic of a conventional wiring approach using coaxial cables, room-temperature microwave generators (AWG), and heavily attenuated coaxial lines connected to the quantum circuit at millikelvin temperatures. The system experiences both passive heat from the thermal conductivity of the coaxial lines and active heat from power dissipation in the attenuators.(b) The cQED metasurface approach, which minimizes hardware overhead by using a single microwave link, multiple DC lines, and a single room-temperature microwave generator, with signals routed directly to the cQED metasurface via an attenuated coaxial line to millikelvin temperatures. The metasurface generates multiple microwave signals at the quantum circuit level with controllable frequency, amplitude, and spatial angle, significantly reducing both passive and active heat loads. (c) Inset: A false-color illustration of a 16-qubit superconducting quantum processor with nearest-neighbour coupling~\cite{ref42}. The lines indicate fixed direct qubit--qubit couplers in a mode where the metasurface functions like an AWG to generate control tones. Alternatively, the metasurface can be configured to mediate the interactions itself through virtual pathways, in which case no fixed direct couplers are required. (d) Inset: The cQED metasurface architecture, consisting of an $N \times N$ square lattice of capacitively shunted Josephson junctions (JJs), coupled to the quantum processor in free space. (e) The equivalent circuit. The metasurface can be designed using JJs or SQUIDs for tunability, with flux loops as shown in the blue circuits added to the diagram.
}
		\label{fig}
	\end{center}
\end{figure*}
These requirements compete against the limited cooling power of the DL, typically around 20 µW. Thus, with current technologies, one could imagine systems with a few thousand qubits at best, far from the threshold necessary for a fault-tolerant universal quantum computer~\cite{ref30}. These critical challenges have motivated new approaches, ranging from building dilution refrigerators with larger cooling capacities, to introducing photonic links that employ optical fibres to guide modulated laser light from room temperature to a cryogenic photodetector for delivering microwave signals at millikelvin temperatures \cite{ref31}. Other efforts focus on developing cryogenic CMOS electronics \cite{ref32}.
While advancements in photonic links and cryogenic CMOS electronics have been significant, issues such as active heating loads, thermal noise, and frequency conversion remain longstanding challenges \cite{ref33, ref34, ref35, ref36}. Another promising approach for cryogenic control is Single Flux Quantum (SFQ) logic, which operates at ultra-fast speeds with low power consumption, making it highly suitable for qubit control. However, SFQ faces challenges, including precision timing control and signal crosstalk, especially in multi-qubit systems\cite{ref37, ref38, ref39, ref40}. Researchers are actively developing solutions, such as advanced circuit designs and shielding techniques, to address these issues, but achieving scalable, high-fidelity control remains an open question \cite{ref41}. Indeed, a low-power cryogenic architecture capable of generating classical signals suitable for high-fidelity qubit operation combined with multiplexing and non-reciprocal functionalities could be instrumental in building a large-scale quantum computer.

In this work, we propose a dynamic metasurface architecture utilising Josephson Junctions (JJs) to generate and multiplex microwave control signals directly at millikelvin temperatures. Our approach consists of a two-dimensional (2D) metasurface composed of an N × N square lattice of capacitively shunted JJs, as illustrated in Fig.~\ref{fig}~(d \& e). This cQED metasurface is designed to be compatible with various quantum circuit architectures, potentially allowing coupling to qubits in free space or on-chip and replacing the heavily attenuated coaxial cables with a single RF coaxial line and multiple DC lines, as shown in Fig.~\ref{fig}~(a \& b), significantly reducing both passive and active heat loads. By modulating the Josephson inductance both spatially and temporally, we enable dynamic control over the frequency, amplitude, phase, and spatial distribution of generated signals.  We derive a nonlinear wave equation governing the system's dynamics, revealing unique phenomena such as beam steering and frequency conversion. Through numerical simulations, we demonstrate the generation of multiple frequency tones with controlled parameters, validating the potential of our approach for efficient, scalable qubit control. Our work lays the foundation for a new paradigm in cryogenic signal generation and routing, potentially overcoming key bottlenecks in quantum computer scalability.

\newpage
\begin{table*}[t!]
    \centering
    \caption{High-level comparison of approaches to address thermal and wiring challenges in superconducting qubit platforms.}
    \label{tab:comparison}
    \begin{tabular}{@{}p{3.5cm}p{3.5cm}p{3.5cm}p{3.5cm}p{3.5cm}@{}}
        \toprule
        Feature & Coax Cables (A) & Cryo-CMOS (B) & Photonic Links (C) & JJ Metasurface (D) (This Work) \\
        \midrule
        Wiring Complexity & Scales linearly with qubit count & Reduced coax but on-chip overhead & Single fiber link + cryogenic detection & Single/few coax lines + DC flux inputs, enabling local generation \\
        Thermal Load at mK & High (many attenuators, conduction) & Limited conduction but transistor dissipation at 4--10 K & Low conduction, complex detection at mK & Potentially low if flux-drive power is small; less conduction from fewer cables \\
        Frequency Generation & Room temperature AWG/LO & On-chip oscillators, limited freq. range & Optical to microwave down-conversion & In-situ JJ nonlinearity for sidebands/harmonics \\
        Scalability & Up to ~100--1000 qubits, wiring-limited & Possibly ~1000+ qubits, still in R\&D & Feasible for moderate scale, integration hurdles & Potentially 1000+ qubits, depending on metasurface array dimensions \& heat budget \\
        Crosstalk/Fidelity & Manageable at small scale, worsens with N & Crosstalk from integrated drivers requires careful design & Loss, device mismatch in conversion & Beam steering and frequency multiplexing can reduce undesired excitations \\
        Speed/Bandwidth & External AWGs, typically up to ~8--10 GHz & Limited by cryo transistor bandwidth & Speed depends on photonic conversion & Potentially high (~ns flux modulation) with wide harmonic coverage \\
        \bottomrule
    \end{tabular}
\end{table*}
Table~\ref{tab:comparison} provides a high-level comparison of four approaches: (A) coax cables with room-temperature electronics, (B) cryo-CMOS, (C) photonic links, and (D) our proposed JJ metasurface. Each method addresses the thermal and wiring trade-offs in unique ways. Coax cables (A) represent a well-established standard but face scalability issues, as each qubit typically requires its own set of lines and attenuators, resulting in a significant wiring burden. Cryo-CMOS (B) reduces the number of external cables by integrating control chips at cryogenic temperatures (4--10 K); however, this introduces challenges such as transistor heat dissipation and noise at millikelvin levels. Photonic links (C) offer the advantage of optical fibers that conduct minimal heat, but the associated cryogenic photodetection and microwave conversion processes add significant complexity. In contrast, the JJ metasurface (D) has the potential to drastically lower cable counts and enable local microwave generation. This advantage hinges on effectively controlling the nonlinearity of JJs with minimal power dissipation, positioning it as a promising approach for scalable quantum control.
While each has merits, (D) the JJ metasurface exploits local parametric modulation of superconducting junctions, potentially addressing thousands of qubits without an exponential cable count.

The thermal load in superconducting qubit systems arises from two primary sources: passive conduction through cables and active dissipation in Josephson Junctions (JJs). Passive heat loads arise from thermal conduction through physical structures such as coaxial cables, mounting hardware, and supports, and can be estimated using Fourier’s law, \( Q_{\text{cond}} = \frac{kA}{L} \Delta T \), where \( k \) is the thermal conductivity, \( A \) is the cross-sectional area, \( L \) is the length, and \( \Delta T \) is the temperature gradient between stages. Active heat loads, by contrast, stem from power dissipation within components inside the cryostat, including resistive elements, RF attenuators, and modulated devices such as Josephson Junctions (JJs). These contribute directly to the mK thermal budget and must be carefully managed to remain within the limited cooling power of dilution refrigerators (typically 10–20 µW at the mK stage) \cite{ref29, ref31}. Quantifying and minimising both contributions is essential for scaling superconducting quantum processors.
Conventional approaches can exceed 100 coaxial lines for even moderate qubit counts, leading to conduction loads of hundreds of microwatts - well above the cooling power of typical dilution refrigerators. Although reducing the number of cables significantly lowers conduction, any JJ-based scheme introduces active dissipation from the Josephson junctions and associated flux pumps, given by
\begin{equation}
P_{\text{JJ,active}} = \sum_{j=1}^N \left[ I_{b,j}^2 R_{\text{jj},j} + P_{\text{mod},j} \right],
\label{eq:PJJactive}
\end{equation}
where \( I_{b,j} \) is the DC bias current applied to junction \( j \), \( R_{\text{jj},j} \) is its normal-state resistance- that is, the effective resistance the junction exhibits when driven out of the superconducting state by strong bias or modulation-, and \( P_{\text{mod},j} \) is the power consumed by the modulation drive circuitry for that junction. Early prototypes, such as \( 8 \times 8 \) and \( 16 \times 16 \) JJ arrays, suggest that the total dissipation at typical flux drive amplitudes remains in the sub-\(\mu\text{W}\) range, well below the conduction limits of modern dilution refrigerators. Scaling to larger arrays will require further validation, but the prospect of drastically reducing cable count often offsets the additional JJ losses, making this metasurface architecture a compelling approach for future quantum processors.


\section{Nonlinear Wave Equation with Time-Space Modulation}
In this section, we derive the nonlinear wave equation governing the dynamics of a capacitively shunted Josephson Junction (JJ) transmission line, as depicted in Fig.~\ref{fig}(d \& e). This derivation lays the groundwork for understanding how modulation of the JJ inductance influences beam control, frequency conversion, and qubit multiplexing. We consider a discrete transmission line composed of identical unit cells of length \( a \). Each unit cell consists of an inductor \( L_J(z, t) \) and a capacitor \( C \), where \( L_J(z, t) \) varies both in space and time due to an applied DC flux \( \Phi_{\text{ext}}(z) \) and an AC modulation signal (RF pump) that modulates the phase difference across the Josephson Junction. The critical current \( I_c(t) \) of the JJ is modulated by an external RF signal, leading to:
\begin{equation}
I_c(t) = I_{c0} \cos\left( \theta(z) + \Delta \varphi \cos(\omega_{\text{RF}} t) \right),
\label{eq:Ic}
\end{equation}
where \( I_{c0} \) is the maximum critical current, \( \theta(z) = \frac{\pi \Phi_{\text{ext}}(z)}{\Phi_0} \) is the static phase shift across the junction due to the applied DC flux \( \Phi_{\text{ext}}(z) \), \( \Delta \varphi \) represents the modulation depth in phase due to the RF pump, \( \omega_{\text{RF}} \) is the frequency of the RF modulation, and \( \Phi_0 = \frac{h}{2e} \) is the magnetic flux quantum. The Josephson inductance \( L_J(z, t) \) is related to the critical current by:
\begin{equation}
L_J(z, t) = \frac{\Phi_0}{2\pi I_c(t)} = \frac{\Phi_0}{2\pi I_{c0}} \frac{1}{\cos\left( \theta(z) + \Delta \varphi \cos(\omega_{\text{RF}} t) \right)}.
\label{eq:LJ}
\end{equation}
To include higher-order harmonics introduced by the modulation, we use the trigonometric identity for the cosine of a sum:
\begin{multline}
\cos\left( \theta(z) + \Delta \varphi \cos\bigl( \omega_{\text{RF}} t \bigr) \right) = \cos\theta(z) \cos\left( \Delta \varphi \cos\bigl( \omega_{\text{RF}} t \bigr) \right) \\
- \sin\theta(z) \sin\left( \Delta \varphi \cos\bigl( \omega_{\text{RF}} t \bigr) \right).
\label{eq:cos_sum}
\end{multline}
Assuming that \( \theta(z) \) is chosen such that \( \sin\theta(z) = 0 \) (i.e., \( \theta(z) = n\pi \) with \( n \in \mathbb{Z} \)), the sine term vanishes, and the expression simplifies to:
\begin{equation}
\cos\left( \theta(z) + \Delta \varphi \cos(\omega_{\text{RF}} t) \right) = \cos\theta(z) \cos\left( \Delta \varphi \cos(\omega_{\text{RF}} t) \right).
\label{eq:cos_simplified}
\end{equation}
Therefore, the Josephson inductance becomes:
\begin{equation}
L_J(z, t) = \frac{\Phi_0}{2\pi I_{c0}} \frac{1}{\cos\theta(z) \cos\left( \Delta \varphi \cos(\omega_{\text{RF}} t) \right)}.
\label{eq:LJ_cos}
\end{equation}
We use the Jacobi-Anger expansion to express \( \cos\left( \Delta \varphi \cos(\omega_{\text{RF}} t) \right) \) as:
\begin{equation}
\cos\left( \Delta \varphi \cos(\omega_{\text{RF}} t) \right) = J_0(\Delta \varphi) + 2 \sum_{n=1}^{\infty} J_{2n}(\Delta \varphi) \cos(2 n \omega_{\text{RF}} t).
\label{eq:jacobi_anger}
\end{equation}
Alternatively, since \( \Delta \varphi \) is small, we can expand the denominator using a Taylor series for \( \cos x \):
\begin{equation}
\cos x = 1 - \frac{x^2}{2} + \frac{x^4}{24} - \cdots.
\label{eq:cos_taylor}
\end{equation}
Substituting \( x = \Delta \varphi \cos(\omega_{\text{RF}} t) \):
\begin{multline}
\cos\left( \Delta \varphi \cos(\omega_{\text{RF}} t) \right) = 1 - \frac{(\Delta \varphi \cos(\omega_{\text{RF}} t))^2}{2} \\+ \frac{(\Delta \varphi \cos(\omega_{\text{RF}} t))^4}{24} - \cdots.
\label{eq:cos_expansion}
\end{multline}
Using trigonometric identities for \( \cos^2(\omega_{\text{RF}} t) \) and \( \cos^4(\omega_{\text{RF}} t) \), we can further expand and obtain:
\begin{multline}
\cos\left( \Delta \varphi \cos(\omega_{\text{RF}} t) \right) = 1 - \frac{(\Delta \varphi)^2}{4} - \frac{(\Delta \varphi)^2}{4} \cos(2 \omega_{\text{RF}} t) \\
+ \frac{(\Delta \varphi)^4}{64} + \frac{(\Delta \varphi)^4}{48} \cos(2 \omega_{\text{RF}} t) \\
+ \frac{(\Delta \varphi)^4}{192} \cos(4 \omega_{\text{RF}} t) - \cdots.
\label{eq:cos_expanded_terms}
\end{multline}
Thus, the denominator becomes:
\begin{multline}
D(t) = \cos\theta(z) \biggl[ 1 - \frac{(\Delta \varphi)^2}{4} - \frac{(\Delta \varphi)^2}{4} \cos(2 \omega_{\text{RF}} t) \\
+ \frac{(\Delta \varphi)^4}{64} + \frac{(\Delta \varphi)^4}{48} \cos(2 \omega_{\text{RF}} t) + \frac{(\Delta \varphi)^4}{192} \cos(4 \omega_{\text{RF}} t) - \cdots \biggr].
\label{eq:D_expansion}
\end{multline}
For small \( \Delta \varphi \), we approximate the reciprocal \( 1/D(t) \) and define \( L_{J0} \) as:
\begin{multline}
 L_{J0} = \frac{\Phi_0}{2\pi I_{c0} D_0} = \frac{\Phi_0}{2\pi I_{c0} \cos\theta(z) \left( 1 - \dfrac{(\Delta \varphi)^2}{4} + \dfrac{(\Delta \varphi)^4}{64} \right)},
\label{eq:LJ0}
\end{multline}
where \( D_0 \) represents the average denominator. Then, the time-dependent modulation of the inductance is:
\begin{equation}
\delta L(z, t) = \frac{(\Delta \varphi)^2}{4 D_0} \cos(2 \omega_{\text{RF}} t).
\label{eq:delta_L}
\end{equation}
The Josephson inductance becomes:
\begin{equation}
 L_J(z, t) = L_{J0} \left( 1 + \delta L(z, t) \right).
\label{eq:LJ_final}
\end{equation}
The wave equation for the voltage \( V(z,t) \) along the transmission line derived from the telegrapher's for \(L(z,t)\) varies slowly with time compared to the signal frequencies of interest and assuming a local relationship:
\begin{multline}
\frac{\partial^2 V(z, t)}{\partial z^2} - L(z, t) C \frac{\partial^2 V(z, t)}{\partial t^2} - C \frac{\partial L(z, t)}{\partial t} \frac{\partial V(z, t)}{\partial t} = 0.
\label{eq:wave_eq}
\end{multline}
where \( C \) is the capacitance per unit length of the transmission line. To address nonlocality, we reformulate the wave equation using a convolutional kernel or Green's function approach. These methods account for memory effects in the system, avoiding unphysical assumptions of instantaneous response. We proceed by updating the wave equation with a convolutional kernel and introducing a Volterra series expansion to capture the nonlinear and nonlocal effects. The kernel-based form of the wave equation that accounts for memory effects is given by:
\begin{equation}
\frac{\partial^2 V(z, t)}{\partial z^2} - C \frac{\partial}{\partial t} \left( \int_{-\infty}^t K(z, t - \tau) \frac{\partial V(z, \tau)}{\partial \tau} \, d\tau \right) = 0,
\label{eq:wave_kernel}
\end{equation}
where \( K(z, t - \tau) \) is the convolutional kernel representing the time-dependent, non-instantaneous response of the inductance. This integral form introduces a causal memory effect, where the voltage response at time \( t \) depends on the history of the voltage over \( \tau < t \), ensuring that changes in inductance are correctly modeled over time. Expressing \( V(z, t) \) as a nonlinear functional series, we have:
\begin{multline}
V(z, t) = \sum_{n=1}^{\infty} \int_{-\infty}^t \cdots \int_{-\infty}^t H_n(z; t - \tau_1, \dots, t - \tau_n) \\
\times \prod_{i=1}^n I(\tau_i) \, d\tau_1 \cdots d\tau_n,
\label{eq:volterra_expansion}
\end{multline}
where \( H_n \) are the Volterra kernels that capture the nonlinearity and nonlocal effects at different orders. Specifically, \( H_1 \) represents the first-order kernel (linear response), while \( H_2 \) and higher-order kernels capture the nonlinear responses. This expansion ensures that each order of nonlinearity is tied to a specific memory function, enabling complex interactions such as harmonic generation and nonlinear coupling to emerge naturally from the system dynamics. If we wish to handle spatial modulation explicitly, we can reformulate the equation using Green's functions. The Green's function \( G(z, z'; t, \tau) \) represents the system's response at \( (z, t) \) due to a unit excitation at \( (z', \tau) \). For a spatially and temporally modulated inductance, the wave equation becomes:
\begin{equation}
V(z, t) = \int_{-\infty}^{\infty} \int_0^{\infty} G(z, z'; \tau) S(z', t - \tau) \, d\tau \, dz'.
\label{eq:greens_function}
\end{equation}
where \( G(z, z'; t - \tau) \) is the Green's function representing the system's impulse response in both space and time, and \( S(z', \tau) \) is the source term, such as an external current or voltage source. This formulation inherently includes both time and spatial memory effects, allowing us to solve for \( V(z, t) \) by convolution with the input source. With these tools, the full wave equation that includes nonlinearity and nonlocality is:
\begin{multline}
\frac{\partial^2 V(z, t)}{\partial z^2} - C \frac{\partial}{\partial t} \left( \int_{-\infty}^t K(z, t - \tau) \frac{\partial V(z, \tau)}{\partial \tau} \, d\tau \right) \\
- \sum_{n=2}^{\infty} \int \cdots \int H_n(z; t - \tau_1, \dots, t - \tau_n) \\
\times \prod_{i=1}^n I(\tau_i) \, d\tau_1 \cdots d\tau_n = 0.
\label{eq:nonlinear_nonlocal_wave_eq}
\end{multline}
In this form, the convolutional term with \( K(z, t - \tau) \) manages the non-instantaneous inductive response, the Volterra series terms \( H_n \) address higher-order nonlinear effects, and spatial nonlocality is inherently accounted for by the Green's function (if needed in an explicit form). The Energy modulation to second order in $\delta L$ is expressed as:
\begin{equation}
P_{\text{mod}}(z,t) = \frac{1}{2} \frac{\partial L_J(z,t)}{\partial t} I^2(z,t)
\label{eq:modulation_power}
\end{equation}
To analyze the wave propagation, we employ coupled-mode theory by expressing \( V(z, t) \) as a sum of harmonics:
\begin{equation}
V(z, t) = \sum_{n=-\infty}^{\infty} V_n(z) e^{-i (\omega + n \omega_{\text{RF}}) t},
\label{eq:V_harmonics}
\end{equation}
where \( \omega \) is the fundamental frequency, and \( V_n(z) \) are the amplitudes of the harmonic components. Substituting and using \( \delta L(z,t) \) from equation (12), we obtain:
\begin{multline}
\sum_{n} \left( \frac{d^2 V_n}{d z^2} + k_n^2 V_n \right) e^{-i (\omega + n \omega_{\text{RF}}) t} + C \frac{\partial L_J(z, t)}{\partial t} \\ \sum_{n} (\omega + n \omega_{\text{RF}}) V_n e^{-i (\omega + n \omega_{\text{RF}}) t} = 0,
\label{eq:coupled_eqs}
\end{multline}
where \( k_n^2 = L_{J0} C (\omega + n \omega_{\text{RF}})^2 \). Collecting terms for each harmonic \( e^{-i(\omega + n \omega_{\text{RF}})t} \), we obtain the coupled equations:
\begin{equation}
\frac{d^2 V_n}{d z^2} + k_n^2 V_n + \sum_{m} M_{n,m} V_m = 0, 
\label{eq:coupled_modes}
\end{equation}
where the coupling matrix \( M_{n,m} \) is given by:
\begin{multline}
M_{n,n+2} = -C \left( \omega + n \omega_{\text{RF}} \right)\\ \left[ \left( \omega + (n + 2) \omega_{\text{RF}} \right) \delta L_{+2} + i \left( \frac{\partial L_J}{\partial t} \right)_{+2} \right], \\
M_{n,n-2} = -C \left( \omega + n \omega_{\text{RF}} \right)\\ \left[ \left( \omega + (n - 2) \omega_{\text{RF}} \right) \delta L_{-2} + i \left( \frac{\partial L_J}{\partial t} \right)_{-2} \right].
\label{eq:M_matrix}
\end{multline}
This modulation introduces coupling between modes with frequencies \( \omega_n \) and \( \omega_{n \pm 2} \), resulting in energy exchange between harmonics and the possibility of frequency conversion. The formation of bandgaps in the dispersion relation implies frequency ranges where wave propagation is prohibited, enabling the design of filters and isolators. Strong modulation requires retaining more terms in the expansion and necessitate numerical methods to solve the eigenvalue problem accurately. Applications include frequency conversion, nonreciprocal devices, parametric amplifiers, and qubit multiplexing in superconducting circuits.
\begin{figure}[!h]
	\begin{center}
		\includegraphics[width=1\columnwidth]{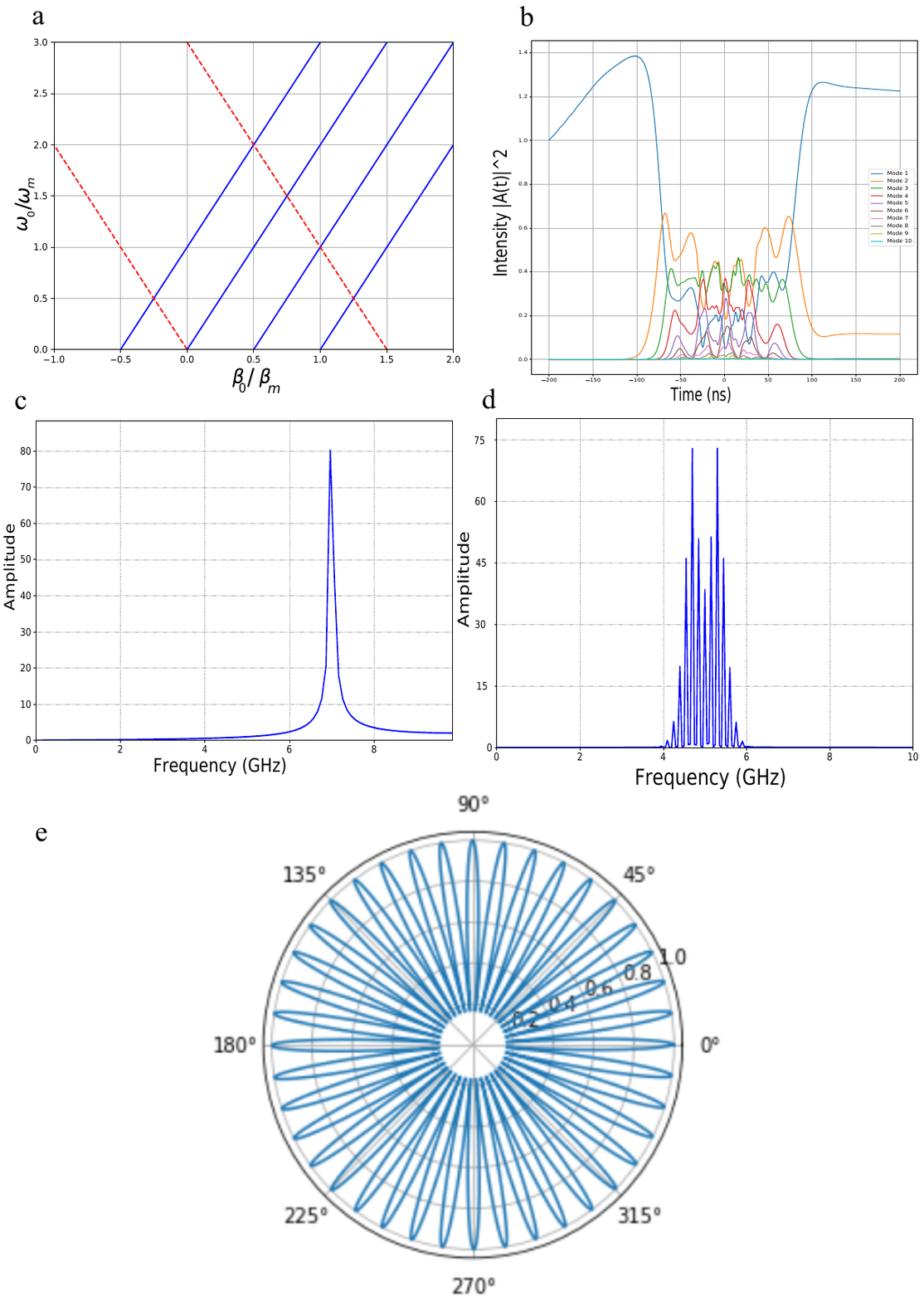}
		\caption{(a) Dispersion diagram for modulation depth $\Delta I = 0.35$, showing asymmetry between forward and backward harmonics, indicative of nonreciprocity and phase matching of all forward harmonics as the system approaches the sonic regime. (b) Time evolution of energy transfer between 10 modes, illustrating how space-time modulation facilitates the distribution of energy from the fundamental mode to higher harmonics. (c) Input frequency spectrum of the fundamental mode. (d) Generated harmonics via frequency up-conversion in the modulated medium. (e) Radiation pattern showing conversion of an isotropic source into 40 directed beams, each addressing a specific qubit in a ring connectivity array, demonstrating precise beam steering and control of diffraction angles.}
		\label{figg}
	\end{center}
\end{figure}

Figure~\ref{figg} provides a comprehensive analysis of the behavior of the space-time modulated JJ transmission line: (a) Dispersion Diagram: Shows the nonreciprocal nature of the medium, with clear asymmetry between forward and backward propagating harmonics, a key feature indicating breaking time-reversal symmetry. This nonreciprocity is crucial for phase matching and controlling harmonic interactions. (b) Time Evolution of Energy Transfer: Demonstrates how energy is redistributed from the fundamental mode to various harmonics over time, facilitated by space-time modulation. (c) Input Frequency Spectrum: Depicts the initial frequency content of the fundamental mode. (d) Generated Harmonics: Illustrates the effectiveness of frequency up-conversion in the modulated medium. (e) Radiation Pattern: Shows the resulting conversion of an initially isotropic source into 40 distinct beams, each precisely directed to a specific qubit in a ring connectivity array, exemplifying the capability of the medium to control beam steering and address individual qubits. These results confirm the ability of the metasurface to perform beam steering and frequency conversion, essential for multiplexed qubit control.
The derived wave equation and beam control mechanisms provide a robust theoretical framework for understanding the dynamic behavior of the metasurface. To further explore and validate these effects, we employ numerical simulations using the Finite-Difference Time-Domain (FDTD) method. The goal of the numerical analysis is to visualize the wave propagation characteristics, investigate the impact of modulation on beam steering, and confirm key concepts such as frequency conversion and nonreciprocity. In the following section, we present the numerical solution to the nonlinear wave equation. These simulations allow us to observe how the modulation parameters influence beam propagation in real time, providing insights into the practical applications of the metasurface in quantum computing. The nonlinearity and dispersion introduced by the modulation result in complex wave interactions, which will be explored in detail through the numerical results.

\subsection{Qubit Interaction Using dynamic metasurface}
Superconducting qubits are commonly controlled using microwave signals transmitted through coplanar waveguides or transmission lines. In our approach, we dynamically engineer the nonlinear wave propagation interactions to facilitate qubit control and entanglement by reconfigurably modulating the Josephson inductance. The time and spatial dependence of the inductance lead to a more complex wave equation, as expected when the series inductance depends on both time and space. Moreover, convolution naturally arises due to the spatial modulation, which must be properly accounted for, even in the linear case. For small modulation depths, \( \Delta_L \), we neglect terms of order \( \Delta_L^2 \). We consider the modulation of the Josephson inductance as:
\begin{equation}
L_J(z, t) = L_{J0} \left[ 1 + \Delta_L \cos(k_m z - \omega_m t) \right],
\label{eq:LJ_modulation}
\end{equation}
where \( L_{J0} \) is the average Josephson inductance, \( \Delta_L \ll 1 \) is the modulation depth, and \( k_m \) and \( \omega_m \) are the modulation wavenumber and frequency, respectively. The flux field \( \Phi(z, t) \) satisfies the wave equation:
\begin{equation}
\frac{\partial}{\partial z} \left( \frac{1}{L_J(z, t)} \frac{\partial \Phi(z, t)}{\partial z} \right) - C_J \frac{\partial^2 \Phi(z, t)}{\partial t^2} = 0,
\label{eq:flux_wave_eq}
\end{equation}
Here \(C_J\) and \(L_{J0}\) denote per-unit-length parameters. Equation~\ref{eq:flux_wave_eq} follows from the Lagrangian density 
\(\mathcal{L}=\tfrac{C_J}{2}\dot\Phi^2-\tfrac{1}{2L_J(z,t)}(\partial_z\Phi)^2\), 
yielding \(C_J\ddot\Phi-\partial_z[(1/L_J)\partial_z\Phi]=0\) even when \(L_J\) is time-dependent.

Assuming the modulation frequency \( \omega_m \) matches the difference between adjacent mode frequencies, i.e., \( \omega_m = \omega_{n+1} - \omega_n \), and considering only nearest-neighbor coupling, we can write:
\begin{multline}
    \ddot{\Phi}_n(t) + \omega_n^2 \Phi_n(t) + \Delta_L \left( \frac{1}{2} \ddot{\Phi}_{n+1}(t) + \frac{k_m}{2 L_{J0} C_J} \Phi_{n+1}(t) \right) e^{-i \omega_m t} \\
    + \Delta_L \left( \frac{1}{2} \ddot{\Phi}_{n-1}(t) - \frac{k_m}{2 L_{J0} C_J} \Phi_{n-1}(t) \right) e^{i \omega_m t} = 0.
    \label{eq:simplified_coupled_eq}
\end{multline}
This set of equations represents the coupled-mode equations, where convolution in space is handled through the mode coupling induced by the modulation. Now, to quantize the system, we expand the flux field \( \Phi(z,t) \) and promote its classical amplitudes to operators. The classical flux field is given by:
\begin{equation}
    \Phi(z,t) = \sum_n \Phi_n(t) e^{i k_n z}.
\end{equation}
Upon quantization, the field becomes:
\begin{equation}
    \hat{\Phi}(z,t) = \sum_n \sqrt{\frac{\hbar}{2\omega_n C_J L}} \left( a_n e^{i k_n z} + a_n^\dagger e^{-i k_n z} \right),
\end{equation}
where \(C_J\) is the capacitance per unit length, \(L\) is the inductance per unit length, and \(a_n\), \(a_n^\dagger\) are the annihilation and creation operators. We fix the mode normalization by enforcing the canonical commutator \([\hat\Phi(z),\hat Q(z')]=i\hbar\delta(z-z')\) with \(\hat Q=C_J\dot{\hat\Phi}\); this gives the prefactor in the field expansion.

The interaction energy arises from the coupling of the qubit to this quantized flux. For a flux qubit, this coupling is proportional to the flux:
\begin{equation}
    H_{int} = \alpha \hat{\sigma}_x \hat{\Phi}(z_q,t),
    \label{eq30}
\end{equation}
where \( \alpha \) is a coupling constant and \( z_q \) is the qubit's position. Expanding this, we have:
\begin{equation}
    H_{int} = \alpha \hat{\sigma}_x \sum_n \sqrt{\frac{\hbar}{2\omega_n C_J L}} \left( a_n e^{i k_n z_q} + a_n^\dagger e^{-i k_n z_q} \right).
\end{equation}
The modulation of the inductance, \( L_J(z,t) = L_{J0} \left[1 + \Delta_L \cos(k_m z - \omega_m t)\right] \), affects the coupling strength, leading to a time-dependent coupling:
\begin{equation}
    \alpha(t) = \alpha_0 \left[1 + \Delta_L \cos(\omega_m t)\right],
\end{equation}
where we have assumed the qubit is located at \(z = 0\) for simplicity. Focusing on a single mode \(n\), the interaction Hamiltonian simplifies to:
\begin{equation}
    H_{int} = \hbar g(t) (\sigma_+ + \sigma_-)(a + a^\dagger),
\end{equation}
where the time-dependent coupling strength is:
\begin{equation}
    g(t) = g_0 \left[1 + \Delta_L \cos(\omega_m t)\right],
\end{equation}
and \( g_0 = \alpha_0 \sqrt{\frac{1}{2\hbar\omega_n C_J L_{J0}}} \). 
Under RWA near \(\omega_{01}\), leakage through the detuned \(|1\rangle\!\leftrightarrow\!|2\rangle\) leg scales as \(p_{\rm leak}\lesssim(\Omega_{12}/\alpha)^2\) with the usual \(\sin^2(\alpha t_g/2)\) envelope; numerics below confirm this bound in our operating range.

\begin{figure}[!t]
	\begin{center}
	\includegraphics[width=1\columnwidth]{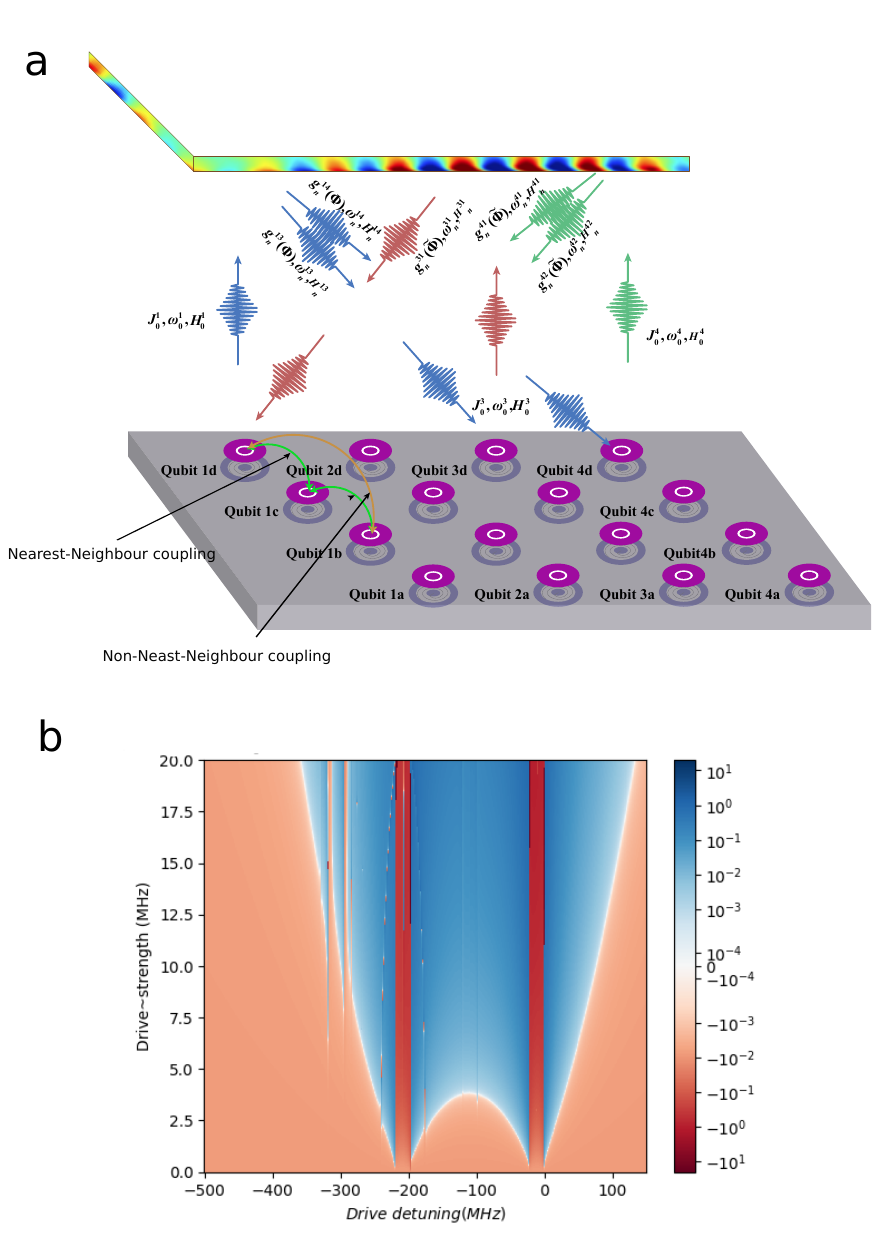}
		\caption{(a) Illustration of a dynamic metasurface enabling two-qubit interactions through both nearest-neighbour and non-nearest-neighbour couplings. The metasurface modulates qubit beam frequencies (e.g. 3-8 GHz), generating harmonics that result in multi-beam diffraction at different angles. Central peaks correspond to the primary incident beams, while side peaks represent higher-order harmonics created by the modulated medium. This setup supports qubit interactions over multiple distance scales, facilitating both next-nearest and non-nearest neighbour couplings. (b) Simulation of the effective ZZ interaction in a two-qubit Sizzle gate mediated by a dynamically modulated metasurface. The color map shows the ZZ interaction strength as a function of drive detuning and drive strength, where darker regions represent stronger interactions. Key parameters include qubit anharmonicities $\alpha_1 = 197$ MHz and $\alpha_2 = 195$ MHz, coupling strength $J = 0.7$ MHz, and detuning $\Delta = 133$ MHz. The modulation facilitates an off-resonant interaction that enables interaction without direct qubit-qubit resonance. The vertical feature near $-300\,\mathrm{MHz}$ corresponds to the $\lvert 1 \rangle \!\to\! \lvert 2 \rangle$ transition (the transmon anharmonicity), confirming multi-level dynamics and leakage channels.}
		\label{figcc}
	\end{center}
\end{figure}

It is worth noting that while Eq.~\ref{eq30} is written in a two-level form for notational clarity, all simulations and effective models retain at least three levels per qubit. In this basis, the loop current operator $\hat{I}_{\ell}$ couples $\lvert 0 \rangle \leftrightarrow \lvert 1 \rangle$ and $\lvert 1 \rangle \leftrightarrow \lvert 2 \rangle$ with matrix elements $I_{01}, I_{12}$, which is essential to capture leakage and ZZ.

\subsection{The Sizzle Gate}
The Sizzle gate is an entangling two-qubit gate that exploits the effective ZZ interaction arising in the dispersive regime of qubit coupling~\cite{ref43}. We consider two qubits located at positions \(z_1\) and \(z_2\), coupled to a dynamic metasurface, such as a modulated nonlinear transmission line. The qubits interact with the metasurface through the Hamiltonian:
\begin{equation}
    H_\text{int} = \hbar g_1 \sigma_x^{(1)} \Phi(z_1, t) + \hbar g_2 \sigma_x^{(2)} \Phi(z_2, t),
\label{eq:qubit_interaction_Hamiltonian}
\end{equation}
where \(g_j\) is the coupling strength of qubit \(j\) to the flux \(\Phi(z,t)\), \(\sigma_x^{(j)}\) is the Pauli-X operator for qubit \(j\), and \(\Phi(z,t)\) includes the effects of the modulation at frequency \(\omega_m\). The dynamic metasurface is modulated in both time and space at a frequency \(\omega_m\), with modulation depth \(\Delta L\), which adjusts the effective coupling between the qubits. The modulation introduces sidebands to the system, resulting in a rich spectrum of coupling pathways without the need to bring the qubits into resonance. Assuming that the qubits are detuned by \(\Delta_{12} = \omega_{q1} - \omega_{q2}\), with \(|\Delta_{12}| \gg g_j\), we operate in the dispersive regime. In this regime, direct exchange of excitations between the qubits is suppressed, and virtual processes dominate the interaction. The effective exchange and ZZ terms used below are obtained from a Schrieffer-Wolff elimination of the multilevel qubit-metasurface Hamiltonian. In particular, the single-qubit dispersive shift is \(\chi\simeq g_{01}^2\alpha/[\Delta(\Delta+\alpha)]\), consequently, ZZ vanishes in the harmonic (\(\alpha\!\to\!0\)) limit and cannot be captured by a strict TLS.

Updating the Hamiltonian to include the time-space modulation, we get:
\begin{multline}
    H_{\text{siZZle, mod}}/h = H_0/h + \sum_{i=0,1} \Omega_i(t, z_i) \cos(2\pi \nu_d t \\+ \phi_i(t, z_i)) (\hat{a}_i^\dagger + \hat{a}_i),
\end{multline}
where \(\Omega_i(t, z_i)\) and \(\phi_i(t, z_i)\) now include modulation effects. Using second-order perturbation theory, we derive an effective Hamiltonian for the qubits that includes a ZZ interaction. The modulated metasurface mediates this interaction through virtual photon exchange. The interaction Hamiltonian can be written as:
\begin{equation}
    H_\text{eff} = \hbar J_\text{eff} \left( \sigma_+^{(1)} \sigma_-^{(2)} + \sigma_-^{(1)} \sigma_+^{(2)} \right),
    \label{eq:effective_Hamiltonian}
\end{equation}
where \(J_\text{eff}\) is the effective exchange coupling between the qubits, modulated by the dynamic metasurface. The effective coupling \( J_{\text{eff}} \) arises due to virtual photon exchange mediated by the modulated modes of the dynamic metasurface. It is influenced by both the time and spatial modulation and is given by:
\begin{equation}
    J_{\text{eff}} = \sum_{n, k} \frac{g_1 g_2 \delta_{n,k} \cos(k_m(z_1 - z_2))}{\Delta_{n,k}},
    \label{eq:J_eff_modulated}
\end{equation}
where \(\delta_{n,k}\) represents the modulation-induced coupling factors for the \(n\)-th mode and the \(k\)-th sideband, \(\Delta_{n,k} = \omega_{q1} - (\omega_n \pm k \omega_m)\) is the effective detuning between qubit 1 and the \(k\)-th sideband of the \(n\)-th mode, and \(k_m\) is the wavevector of the modulation. The cosine term reflects the spatial dependence of the interaction due to the qubits' positions \(z_1\) and \(z_2\). The sum is over the modes and sidebands that contribute to the interaction. If a dominant mode and sideband combination is found, the expression simplifies to:
\begin{equation}
    J_{\text{eff}} \approx \frac{g_1 g_2 \delta \cos(k_m (z_1 - z_2))}{\Delta},
    \label{eq:J_eff_simplified_modulated}
\end{equation}
where \(\delta\) is the effective coupling factor for the dominant mode and sideband, and \(\Delta = \omega_{q1} - (\omega_n \pm k \omega_m)\) is the corresponding detuning. Incorporating the time-space modulation into the ZZ interaction, the effective Hamiltonian becomes:
\begin{multline}
    H_{\text{eff}}/h = \tilde{\nu}_{ZZ, \text{mod}} \sigma_z \otimes \sigma_z / 4 \\
    + \tilde{\nu}_{IZ, \text{mod}} \sigma_z \otimes I / 4 + \tilde{\nu}_{ZI, \text{mod}} I \otimes \sigma_z / 4,
\end{multline}
where \(\tilde{\nu}_{ZZ, \text{mod}}\), \(\tilde{\nu}_{IZ, \text{mod}}\), and \(\tilde{\nu}_{ZI, \text{mod}}\) now include modulation effects:
\begin{equation}
    \tilde{\nu}_{ZZ, \text{mod}} = \tilde{\nu}_{ZZ} + \frac{2 J \alpha_0 \alpha_1 \Omega_0 \Omega_1 \cos(\phi_0 - \phi_1)}{(\Delta_{0,d} + \alpha_0)(\Delta_{1,d} + \alpha_1) \Delta_{0,d} \Delta_{1,d}},
\end{equation}
with \(\Omega_i\) and \(\phi_i\) as functions of time and space.
This Hamiltonian describes a ZZ interaction between the qubits, where the energy of each qubit depends on the state of the other qubit. The modulation parameters, such as the modulation frequency \(\omega_m\), depth \(\Delta L\), and the wavevector \(k_m\), control the effective coupling between the qubits. The gate time \(t_g\) for the Sizzle gate is determined by the accumulated phase due to the ZZ interaction. To achieve a desired controlled-Z (CZ) gate with phase \(\theta\), we set the interaction time such that:
\begin{equation}
    \left( \frac{(J_\text{eff})^2}{\Delta_{12}} \right) t_g = \theta,
    \label{eq:gate_time_modulated}
\end{equation}
For a maximally entangling gate, such as a controlled-Z (CZ) gate where \(\theta = \pi\), the interaction time is:
\begin{equation}
    t_g = \frac{\pi \Delta_{12}}{(J_\text{eff})^2}.
    \label{eq:CZ_gate_time_modulated}
\end{equation}
This scaling follows from \(J_{\rm eff}\)-mediated virtual exchange in the \(|J_{\rm eff}|\ll|\Delta_{12}|\) dispersive limit; constant prefactors depend on pulse shaping and detuning symmetry and are verified numerically in Fig.~\ref{figcc}(b).

A prime example is the Sizzle gate, exploiting sidebands near each qubit frequency 
\(\omega_{q1}\) and \(\omega_{q2}\). By pumping the metasurface at 
\(\omega_{m} \approx |\omega_{q1} - \omega_{q2}|\), one can create bridging sidebands 
that yield an effective Hamiltonian:
\begin{equation}
H_{\mathrm{eff}} \;=\; \hbar \sum_{n, k} 
{\frac{g_{1}\,g_{2}\,\delta_{n,k}\,\cos\!\bigl[k_{m}\,(z_{1}-z_{2})\bigr]}{\Delta_{n,k}}}\,
\Bigl(\sigma_{+}^{(1)}\,\sigma_{-}^{(2)} + \mathrm{h.c.}\Bigr).
\label{eq:Sizzle}
\end{equation}

Note that the effective exchange and ZZ interactions follow from a Schrieffer--Wolff elimination of the multilevel Hamiltonian. In practice, the ZZ term arises from virtual processes $\lvert 1 \rangle \rightarrow \lvert 2 \rangle$, which scale as $\chi \simeq g_{01}^2 \alpha / \big[\Delta (\Delta + \alpha)\big]$. This dependence vanishes for $\alpha \rightarrow 0$, illustrating why a multilevel description is required. Our simulations (Fig.~\ref{figcc}(b)) explicitly show the leakage resonance near $-300~\text{MHz}$, directly corresponding to the $\lvert 1 \rangle \rightarrow \lvert 2 \rangle$ transition and associated anharmonicity leakage channels.

\subsection{Multiplexing Capacity and Gate Fidelity Analysis}
To evaluate the practical scalability of our space-time modulated Josephson junction (JJ) metasurface, we now present a quantitative design example, including spatial resolution, frequency channel separation, and thermal load at the dilution refrigerator stage. Our goal is to determine how many qubits can be reliably controlled using a single modulated microwave source and a programmable JJ metasurface.

First, we evaluate the spatial resolution of the metasurface. We consider a metasurface formed by an \( N_{\text{JJ}} \times N_{\text{JJ}} \) square array of JJs with lattice constant \( a = 10\,\mu\text{m} \). For \( N_{\text{JJ}} = 100 \), the metasurface has a diameter \( D = N_{\text{JJ}} a = 1\,\text{mm} \). This aperture is sub-wavelength at 5\,GHz, corresponding to free-space wavelength is \( \lambda_0 = 2\pi c / \omega_0 \approx 60\,\text{mm} \). Hence, the far-field diffraction formulas \( \frac{\pi D^2/4}{(\lambda_0/2)^2} \) do not directly apply. Instead, qubits are coupled in the near field of the metasurface, allowing localized control fields. In practice, the number of qubits that can be simultaneously addressed is determined by the number of independent modulation segments and the achievable isolation between focal regions, rather than by diffraction-limited beam count. For example, if the metasurface is partitioned into \( M \) groups of cells (each driven by a separate DC flux bias line), up to \( M \) distinct beams or near-field modes can be generated at once. In our example, we conservatively target controlling on the order of tens of qubits in parallel. In addition to spatial segmentation, the metasurface supports frequency-division multiplexing across a wide modulation bandwidth. Each group of JJ cells can be driven with RF pulses containing one or more carefully chosen tone frequencies to address different qubits. The JJ metasurface’s nonlinearity creates sidebands around the 5\,GHz carrier, effectively shifting microwave energy into tunable frequency channels. We assume a modulation frequency up to \( \omega_s/2\pi = 2\,\text{GHz} \), which sets the available spectrum for sidebands (approximately 3–7\,GHz around the carrier). To avoid spectral overlap between control tones, a minimum channel spacing \( \Delta f_{\min} \approx 50\text{--}100\,\text{MHz} \) is required -- on the order of typical transmon linewidths (few MHz) with a guard band for detuning tolerance. The maximum number of frequency channels is then roughly given by the ratio of modulation bandwidth to channel spacing. For example, using \( \Delta f_{\min} = 50\,\text{MHz} \) and \( \omega_s/2\pi = 2\,\text{GHz} \) yields about:
\begin{equation}
N_{freq} = \frac{\omega_s / 2\pi}{\Delta f_{\min}} \approx 40.
\end{equation}
Additionally, to avoid collisions with 
$\lvert 1 \rangle \!\to\! \lvert 2 \rangle$ transitions we require 
$\lvert f_j - f_k - |\alpha_k| \rvert \gtrsim 60$--$100\,\mathrm{MHz}$; 
the leakage feature near $-300\,\mathrm{MHz}$ in Fig.~\ref{figcc}(b) 
illustrates this constraint. This ensures each frequency channel lies outside the typical transmon linewidth.

Second, we examine gate fidelity under frequency-multiplexed operation. The total gate infidelity for qubit \( j \) includes multiple contributions:
\begin{equation}
1 - F_j = \epsilon_{\text{XT}}^{(j)} + \epsilon_{\text{dec}}^{(j)} + \epsilon_{\text{amp}}^{(j)} + \epsilon_{\text{phase}}^{(j)},
\end{equation}
where \( F_j \) denotes the overall fidelity of the quantum gate applied to qubit \( j \). The term \( \epsilon_{\text{XT}}^{(j)} \) represents the infidelity arising from crosstalk with other frequency channels, due to off-resonant driving by harmonics not intended for qubit \( j \). The term \( \epsilon_{\text{dec}}^{(j)} \) accounts for decoherence during the gate time, including contributions from both energy relaxation and dephasing. The contribution \( \epsilon_{\text{amp}}^{(j)} \) quantifies errors due to imperfections in the amplitude of the drive pulse, which may result from control electronics or calibration drift. Finally, \( \epsilon_{\text{phase}}^{(j)} \) captures infidelity due to phase noise, including timing jitter and oscillator instability, which lead to coherent phase errors during the gate. \( \epsilon_{\text{XT}}^{(j)} \) infidelity due to off-resonant coupling to other frequency tones is given by:
\begin{equation}
\epsilon_{\text{XT}}^{(j)} = \sum_{n \neq j} \left|\frac{g_{jn}}{g_{jj}}\right|^2 \frac{\Omega_R^2}{(\omega_j - \omega_n)^2} C_{\text{NL}}(n,j).
\end{equation}
Here, \( \Omega_R \) is the Rabi frequency, and \( C_{\text{NL}}(n,j) \) is a nonlinear spectral suppression factor:
\begin{equation}
C_{\text{NL}}(n,j) = \frac{|J_{|n-j|}(\beta)|^2}{|J_0(\beta)|^2} \cdot \frac{1}{1 + \beta^2(n-j)^2/4},
\end{equation}
with \( \beta = \pi \Phi_{\text{rf}} / \Phi_0 \) the normalized modulation amplitude. The first fraction is the relative spectral weight of the unwanted sideband (order $|n-j|$) compared to the carrier, and the second factor phenomenologically accounts for an increasing suppression of far-detuned tones (for large $|n-j|$) by circuit bandwidth limits. Through careful engineering, we choose $\beta$ and incorporate pulse shaping and filters within the metasurface such that $C_{\rm NL}\ll 1$ for all $n\neq j$, resulting in unwanted tone amplitudes being suppressed by $>60$ dB, effectively limiting $\epsilon_{XT}$ to the $10^{-6}$–$10^{-5}$ range. The cumulative crosstalk error becomes $\epsilon_{\text{XT,total}} \approx 39 \times \epsilon_{\text{XT,single}} \approx 3.9 \times 10^{-5}$ to $3.9 \times 10^{-4}$. With isolation of $>66$ dB and coherence times baseline $T_1=T_2=300~\mu\text{s}$, we reach gate errors $\epsilon_{\rm dec}\sim(0.4\text{--}10)\times 10^{-4}$ for $t_g=25\text{--}200$ ns. Overall, high-fidelity gates are achievable for 40 multiplexed tones, with $F_{1Q}\gtrsim 0.999$ and $F_{2Q}\sim0.999\text{--}0.9996$. In a roadmap scenario with $T_1\approx 4$ ms, the decoherence contribution would be an order of magnitude smaller.

Third, we evaluate achievable Rabi frequencies and gate durations. From FDTD simulations, the electric field at the qubit plane reaches \( |\mathcal{E}_j| \approx 0.1{-}0.5\,\text{V/m} \) for $1~\mu\text{W}$ input power. The Rabi frequency is:
\begin{equation}
\Omega_R^{(j)} = \frac{2 \mu_j}{\hbar} |\mathcal{E}_j|, \quad \text{yielding } \Omega_R / 2\pi \approx 10{-}20\,\text{MHz},
\end{equation}
and \( \pi \)-rotation gate times of:
\begin{equation}
t_{\pi} = \frac{\pi}{\Omega_R} \approx 25{-}50\,\text{ns}.
\end{equation}
A controlled-Z gate implemented via Stark-shift detuning has a time:
\begin{equation}
t_{\text{CZ}} = \frac{\pi \Delta}{4g^2} \approx 50{-}200\,\text{ns},
\end{equation}
for \( g/2\pi \sim 10{-}30~\text{MHz} \) and \( \Delta/2\pi \sim 100~\text{MHz} \).
Finally, we estimate the thermal budget. Total dissipation includes:
\begin{equation}
P_{\text{total}} = P_{\text{static}} + P_{\text{dynamic}} + N_{\text{qubit}} P_{\text{per-qubit}}.
\end{equation}
The static power dissipation is:
\begin{equation}
P_{\text{static}} = N_{\text{JJ}}^2 I_{\text{bias}}^2 R_n \approx 1\,\mu\text{W},
\end{equation}
assuming \( I_{\text{bias}} = 0.1 I_c \), \( R_n = 100\,\Omega \).
The dynamic RF losses are:
\begin{equation}
P_{\text{dynamic}} = \sum_{m,n} \frac{\omega_m C_J V_{\text{rf}}^2}{2Q_{mn}} \approx 3\,\mu\text{W}.
\end{equation}
The power delivered to each qubit is:
\begin{equation}
P_{\text{per-qubit}} = \eta_{\text{coup}} P_{\text{in}} |J_n(\beta)|^2 L_{\text{path}}^{-1} \approx 0.1\,\mu\text{W}.
\end{equation}
For \( N_{\text{qubit}} = 45 \), this yields a total power budget of:
\begin{equation}
P_{\text{total}} = 1 + 3 + 4.5 = 8.5\,\mu\text{W},
\end{equation}
which is well within the $20\,\mu\text{W}$ cooling power capacity of typical dilution refrigerators. To ensure high-fidelity operation with 45 qubits, the required system-level specifications include frequency spacing of at least $50$--$100\,\text{MHz}$ between adjacent channels, spatial isolation greater than $30\,\text{dB}$ between addressed sites, suppression of unwanted harmonics by more than $40\,\text{dB}$, and phase noise below $10\,\text{mrad}$ over the duration of a single-qubit gate. These specifications are within reach of current nanofabrication and microwave engineering technologies. With $T_1=T_2=300~\mu$s, the architecture supports simultaneous, high-fidelity control of $N_{\text{qubit}} = 40$--45 qubits with gate speeds comparable to state-of-the-art microwave control. With longer coherence times approaching $4$ ms, fidelities would further improve by about an order of magnitude.
\begin{table}[t]
\centering
\caption{Key Parameters for Multiplexed Qubit Control}
\begin{tabular}{lcc}
\hline
\textbf{Parameter} & \textbf{Symbol} & \textbf{Value} \\
\hline
Number of JJs per side & \( N_{\text{JJ}} \) & 100 \\
Unit cell size & \( a \) & \( 10\,\mu\text{m} \) \\
Metasurface diameter & \( D \) & \( 1\,\text{mm} \) \\
Operating frequency & \( \omega_0/2\pi \) & \( 5\,\text{GHz} \) \\
Modulation frequency & \( \omega_s/2\pi \) & \( 2\,\text{GHz} \) \\
Maximum frequency channels & \( N_{\text{freq}} \) & 40 \\
Channel spacing & \( \Delta f_{\min} \) & \( 50\,\text{MHz} \) \\
Critical current & \( I_c \) & \( 1\text{--}10\,\mu\text{A} \) \\
Junction capacitance & \( C_J \) & \( 10\text{--}100\,\text{fF} \) \\
Input power & \( P_{\text{in}} \) & \( 1\,\mu\text{W} \) \\
Gate time (single-qubit) & \( t_{\pi} \) & \( 25\text{--}50\,\text{ns} \) \\
Required isolation ($F\gtrsim 0.999$) & -- & $60$--$66\,\text{dB}$ \\
Required coherence time (baseline) & \( T_1,T_2 \) & $300\,\mu\text{s}$ \\
Number of qubits & \( N_{\text{qubit}} \) & 40\text{--}45 \\
Total power consumption & \( P_{\text{total}} \) & \( 8.5\,\mu\text{W} \) \\
Cooling power budget & \( P_{\text{cool}} \) & \( 20\,\mu\text{W} \) \\
\hline
\end{tabular}
\label{tab:parameters}
\end{table}

\section{Numerical Solution of the Nonlinear Wave Equation with Time-Space Modulation}

We solve the nonlinear wave equation (Eq.~\eqref{eq:nonlinear_nonlocal_wave_eq}) using the Finite-Difference Time-Domain (FDTD) method. This method discretizes both space and time, representing \( \phi(z, t) \) as \( \phi_{i}^{n} \), where \( i \) and \( n \) are spatial and temporal grid indices, respectively:
\[
\phi(z, t) \rightarrow \phi_{i}^{n}, \quad z = i \Delta z, \quad t = n \Delta t.
\]
We approximate the derivatives using central differences. For time derivatives:
\begin{multline}
    \frac{\partial \phi}{\partial t} \Big|_{i}^{n} \approx \frac{\phi_{i}^{n+1} - \phi_{i}^{n-1}}{2 \Delta t}, \quad 
    \frac{\partial^2 \phi}{\partial t^2} \Big|_{i}^{n} \approx \frac{\phi_{i}^{n+1} - 2\phi_{i}^{n} + \phi_{i}^{n-1}}{(\Delta t)^2},
\end{multline}
For spatial derivatives:
\begin{multline}
    \frac{\partial \phi}{\partial z} \Big|_{i}^{n} \approx \frac{\phi_{i+1}^{n} - \phi_{i-1}^{n}}{2 \Delta z}, \quad 
    \frac{\partial^2 \phi}{\partial z^2} \Big|_{i}^{n} \approx \frac{\phi_{i+1}^{n} - 2\phi_{i}^{n} + \phi_{i-1}^{n}}{(\Delta z)^2}.
\end{multline}
Higher-order mixed derivatives are given by:
\begin{multline}
    \frac{\partial^4 \phi}{\partial z^2 \partial t^2} \Big|_{i}^{n} \approx \frac{1}{(\Delta z)^2 (\Delta t)^2} \left( \phi_{i+1}^{n+1} - 2\phi_{i}^{n+1} + \phi_{i-1}^{n+1} \right. \\
    \left. - 2\left( \phi_{i+1}^{n} - 2\phi_{i}^{n} + \phi_{i-1}^{n} \right) + \phi_{i+1}^{n-1} - 2\phi_{i}^{n-1} + \phi_{i-1}^{n-1} \right).
\end{multline}
Substituting the central difference approximations into the wave equation, we obtain the fully discrete equation:
\begin{multline}
    C \left( \frac{\phi_{i}^{n+1} - 2\phi_{i}^{n} + \phi_{i}^{n-1}}{(\Delta t)^2} \right) \\
    - C_J a^2 \Bigg( \frac{\phi_{i+1}^{n+1} - 2\phi_{i}^{n+1} + \phi_{i-1}^{n+1}}{(\Delta z)^2 (\Delta t)^2} 
    - 2\left( \phi_{i+1}^{n} - 2\phi_{i}^{n} + \phi_{i-1}^{n} \right) \\
    + \phi_{i+1}^{n-1} - 2\phi_{i}^{n-1} + \phi_{i-1}^{n-1} \Bigg) \\
    + \frac{I_{c0} \Delta_I k_m}{\varphi_0} a \sin(k_m z_i - \omega_m t_n) \sin\phi_{i}^{n} \\
    - \frac{I_c(z_i, t_n)}{\varphi_0} \left( a \cos\phi_{i}^{n} \frac{\phi_{i+1}^{n} - \phi_{i-1}^{n}}{2 \Delta z} \right. \\
    \left. - a^2 \sin\phi_{i}^{n} \left( \frac{\phi_{i+1}^{n} - \phi_{i-1}^{n}}{2 \Delta z} \right)^2 \right) = 0.
\end{multline}
This equation allows us to solve for \( \phi_i^{n+1} \) in terms of previously known values at time step \( n \). By discretizing both time and space, we represent the field 
\( \phi_(z,t) \) at discrete points on a grid, allowing us to compute wave propagation in a modulated transmission line. The central difference approximations for time and space derivatives are applied to convert the continuous wave equation into a fully discrete form, which is iterative solved to model wave evolution over time.
\begin{figure}[!t]
	\begin{center}
		\includegraphics[width=0.9\columnwidth]{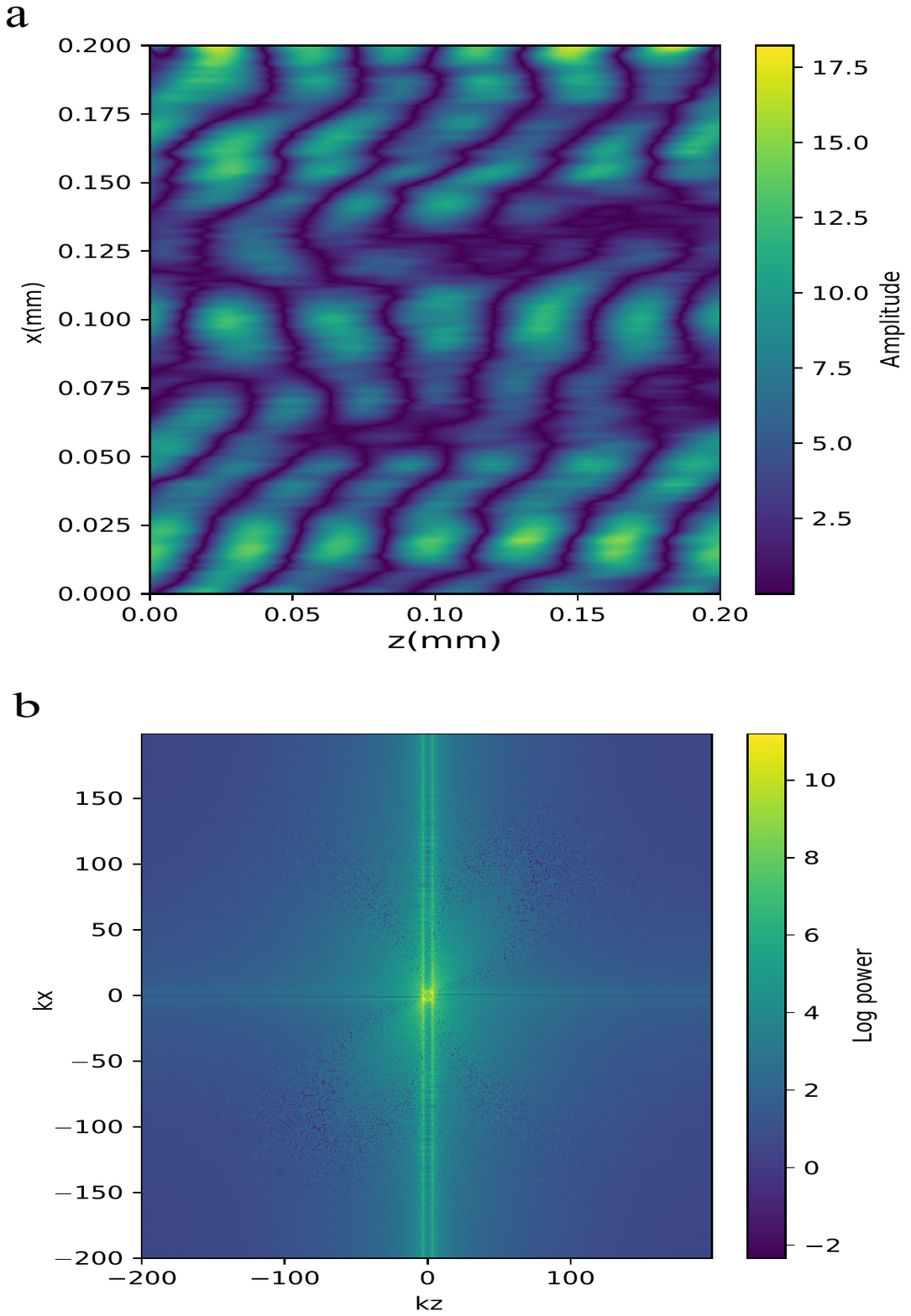}
		\caption{(a) Spatial amplitude distribution in the time-space modulated Josephson Junction transmission line showing wave interference in the $x-z$ plane. (b) Illustration of the conversion of the incident wave to surface waves along $k_z$ and $k_x$, with multiple beams formed at different angles.
    }
		\label{fig1}
	\end{center}
\end{figure}
We begin by setting the initial conditions for \( \phi_i^0 \) and \( \phi_i^1 \), which represent the field at the first two time steps. Depending on the physical scenario, this could be a Gaussian pulse or sinusoidal wave. Additionally, the initial time derivative \( \frac{\partial \phi}{\partial t} \Big|_{t=0} \) is needed to compute \( \phi_i^1 \). Appropriate boundary conditions are applied at the edges of the computational domain. For open boundaries, we use Mur’s first-order absorbing boundary conditions to minimize reflections, allowing waves to propagate out of the domain without artificial reflections. Alternatively, periodic or fixed boundary conditions may be applied depending on the specific problem. The time-stepping process involves updating the field \( \phi_{i}^{n+1} \) at each grid point \( i \) for each time step \( n \), using the fully discretized equation. The nonlinear terms, such as \( \sin\phi_{i}^{n} \), \( \cos\phi_{i}^{n} \), and spatial derivatives \( \left( \frac{\phi_{i+1}^{n} - \phi_{i-1}^{n}}{2 \Delta z} \right)^2 \), are computed at the current time step. To ensure numerical stability, we select \( \Delta t \) and \( \Delta z \) based on the Courant-Friedrichs-Lewy (CFL) condition \( \Delta t \leq \Delta z / v_{\text{max}} \) and \(
S = \frac{c \Delta t}{\sqrt{(\Delta x)^2 + (\Delta z)^2}} \leq \frac{1}{\sqrt{2}}.\) However, due to nonlinearity and dispersion, a more stringent condition may be required, which can be determined through von Neumann stability analysis. Relevant quantities, such as \( \phi_{i}^{n} \), \( \frac{\partial \phi_{i}^{n}}{\partial t} \), and other derived fields, are recorded for the analysis of phenomena like frequency conversion, parametric amplification, and dispersion characteristics.

\subsection{Simulation Results}
\begin{figure}[!t]
	\begin{center}
		\includegraphics[width=1\columnwidth]{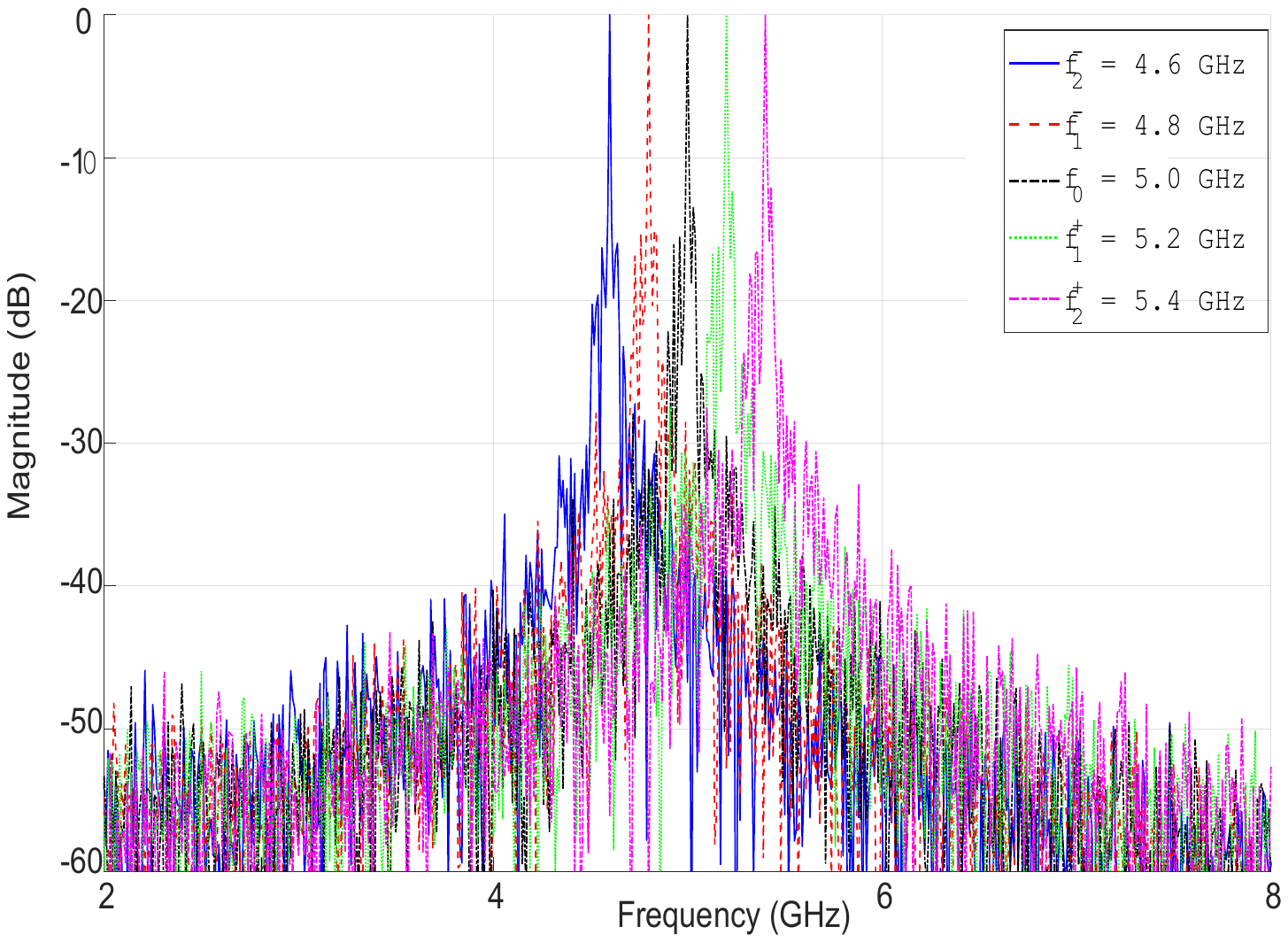}
		\caption{Frequency spectra of incident and diffracted beams for various input frequencies (4.6-5.4 GHz). Central peaks represent incident beams, while side peaks show harmonics generated by the modulated medium, illustrating multi-beam formation at different angles. The amplitude is normalized for clarity of comparison across different frequency ranges.
    }
		\label{fig2}
	\end{center}
\end{figure}
We present the FDTD simulation results for wave propagation in a space-time modulated superconducting metasurface. The incident wave is modulated with a Gaussian envelop and an incident wave frequency of 5 GHz. The spatial grid resolution is set at \( \lambda_0 / 25 \) with time step determined by the Courant condition. Mur boundary conditions are applied to minimize reflections. Figure~\ref{fig1} illustrates the spatial and spectral characteristics of the wave propagation in a time-space modulated Josephson Junction transmission line. In (a), the spatial distribution of the wave amplitude is shown in the \(x-z\) plane. The interference pattern is a result of the modulation, with distinct regions of high (yellow) and low (purple) amplitudes. Figure~\ref{fig1} (b) presents an illustration of the spatial pattern, revealing the incident wave propagation out of the plane. Notably, the wave is converted into surface waves along the and \(k_z\) and \(k_x\) directions, with the formation of multiple beams at different angles. These beams correspond to energy redistribution into various harmonic modes, as shown by the bright spots in the logarithmic power spectrum.

Figure \ref{fig2} Frequency spectrum analysis of multi-beam formation in a modulated medium. The Fourier transform illustrates the frequency content of the incident beam (central peak) and the generation of multiple harmonics (side peaks) corresponding to diffracted beams at various angles. Each color represents a different fundamental frequency ($f_{0}$) of the incident wave, ranging from 4.6 GHz to 5.4 GHz. The plot demonstrates how the modulation of the medium leads to frequency mixing, resulting in the creation of new frequency components and the spatial separation of beams. To confirm that the crosstalk remains negligible under this configuration, we evaluate Eq.~(47) using simulated parameters. From Fig.~5, the nonlinear spectral suppression factor is $C_{\rm NL}(n,j) \approx 5 \times 10^{-4}$, and we estimate the relative off-target coupling ratio as $g_{jn}/g_{jj} \approx 0.022$. For a Rabi frequency $\Omega_R / 2\pi = 50$\,MHz and a qubit frequency spacing of $\Delta = 100$\,MHz, the crosstalk error per-gate fidelity is $F_j \approx 0.9999998$ with respect to crosstalk alone—well above the threshold for fault-tolerant operation.

\section{Discussion and Outlook}
This work underscores the potential of space-time modulated Josephson Junction (JJ) metasurfaces in achieving scalable, hardware-efficient control of superconducting qubits. By harnessing the nonlinear dynamics and modulation capabilities inherent in JJ arrays, the proposed architecture addresses significant challenges associated with scaling quantum processors. Notably, it mitigates wiring complexity and reduces thermal loads-critical factors when scaling to large qubit numbers. The ability to multiplex control signals through beam steering and frequency conversion allows for the simultaneous operation of multiple qubits using a single input, significantly simplifying the control infrastructure. The metasurface architecture also enables the multiplexing of control signals to multiple qubits by modulating different regions to generate pulses with varying frequencies or phases, minimizing hardware overhead.

Challenges remain in achieving the necessary modulation depth and frequency precision to avoid errors that affect gate fidelity. Material choices for JJs impact performance~\cite{ref45}, with materials such as NbN/AlN/NbN~\cite{ref44} offering stronger modulation and higher signal amplitudes but introducing fabrication complexities and potential microwave losses. Thermal management is another critical factor. While the metasurface reduces the number of control lines, the modulation process itself can introduce thermal loads that must be managed within the cooling capacity of dilution refrigerators. Crosstalk and interference between control signals must be mitigated using strategies such as spatial filtering, frequency multiplexing, and advanced modulation schemes.

Future work will concentrate on experimental prototyping of the metasurface, optimizing modulation schemes, addressing thermal management, and exploring adaptive modulation driven by real-time feedback. Implementing error mitigation strategies and studying scalability will be critical for evaluating the metasurface’s performance as qubit counts increase. Beyond control multiplexing, recent theoretical work has shown that Josephson junction metasurfaces can also mediate long-range entangling interactions between distant qubits, offering a potential architecture for scalable quantum interconnects~\cite{ref46}. Investigating this complementary functionality will be crucial for assessing the metasurface’s dual role in both control and entangling operations. The successful execution of quantum algorithms using metasurface-controlled qubits will be a key milestone, demonstrating the practical benefits of this approach. In conclusion, space-time modulated JJ metasurfaces offer a promising path towards scalable and hardware-efficient qubit control. Interdisciplinary collaboration across superconducting electronics, quantum information science, and materials engineering will be essential to realizing the full potential of this approach and accelerating the development of large-scale quantum processors.

\begin{acknowledgments}
M. Bakr acknowledges support from EPSRC QT Fellowship Grant
EP/W027992/1 and EP/Z53318X/1. We thank Smain Amari and Mohammed Alghadeer for insightful discussions.
\end{acknowledgments}


\end{document}